\documentclass[%
reprint,
 amsmath,amssymb,
 aps,
]{revtex4-1}

\usepackage{xcolor}
\usepackage{graphicx}
\usepackage{dcolumn}
\usepackage{bm}
\usepackage{hyperref}
\usepackage{amssymb}
\usepackage{physics}
\newcommand{\figref}[1]{FIG.\ref{#1}}

\begin{document}


\title{Calibration of the cross-resonance two-qubit gate between directly-coupled transmons}
\author{A.~D.~Patterson}
 \altaffiliation[Now at ]{Oxford Quantum Circuits, King Charles House, Oxford. OX1 1JD}
 \email{apatterson@oxfordquantumcircuits.com}
\author{J.~Rahamim}
\author{T.~Tsunoda}
\author{P.~Spring}
\author{S.~Jebari}
\author{K.~Ratter}
\author{M.~Mergenthaler}
\author{G.~Tancredi}
\author{B.~Vlastakis}
\author{M.~Esposito}
\author{P.~J.~Leek}
\affiliation{%
 Clarendon Laboratory, Department of Physics, University of Oxford, OX1 3PU, Oxford, United Kingdom}%

\date{\today}

\begin{abstract}
Quantum computation requires the precise control of the evolution of a quantum system, typically through application of discrete quantum logic gates on a set of qubits. Here, we use the cross-resonance interaction to implement a gate between two superconducting transmon qubits with a direct static dispersive coupling. We demonstrate a practical calibration procedure for the optimization of the gate, combining continuous and repeated-gate Hamiltonian tomography with step-wise reduction of dominant two-qubit coherent errors through mapping to microwave control parameters. We show experimentally that this procedure can enable a $\hat{ZX}_{-\pi/2}$ gate with a fidelity $F=97.0(7)\%$, measured with interleaved randomized benchmarking. We show this in a architecture with out-of-plane control and readout that is readily extensible to larger scale quantum circuits.

\end{abstract}

\pacs{Valid PACS appear here}

\maketitle


\section{\label{sec:introduction}Introduction}
A variety of hardware platforms are currently under intense development towards the realisation of useful quantum computers, with several reaching a maturity at which moderate fidelity quantum control can now be routinely achieved in few-qubit systems \cite{Debnath2016, Lieaar3960, PhysRevX.6.021040, PhysRevLett.117.210502, Neill195}. A key requirement for scaling such platforms to a practically useful level of quantum computation is the precise calibration of two-qubit quantum logic gates to high fidelity, in an architecture that is practically scalable to many qubits \cite{Knill2005, Fowler12}. This applies both for the development of useful imperfect near-term devices \cite{Preskill2018}, as well for the pursuit of fully fault-tolerant error-corrected universal machines \cite{DiVincenzo96, Campbell2017}. In the superconducting circuit platform, a large variety of methods for implementing two-qubit quantum logic have been proposed and demonstrated \cite{Gambetta2017}, in all cases using precisely-shaped analog microwave pulses to deliver the control via microwave transmission lines. Coherent errors associated with imperfections in this microwave control can be mitigated through precise calibration of the analog control pulse parameters.

In a superconducting circuit architecture employing fixed-frequency, statically dispersively-coupled qubits, the microwave-activated cross-resonance interaction \cite{Rigetti2010} can be used to implement two-qubit entangling operations \cite{Chow2011, Sheldon2016}. The interaction is enabled by the application of a cross-resonant drive tone; a qubit drive which is resonant with the transition of a neighbouring qubit. As this tone can be applied to the device using the same control wiring as for single qubit operations, cross-resonance allows two-qubit gate implementation with little adaptation to processor design or control circuitry. This interaction has been successfully implemented through mutual coupling to an interaction-mediating resonator \cite{Chow2011}, allowing for the qubits to be well spatially separated, which in principle keeps cross-talk between the qubits and their associated control wiring low. In practice, control errors due to imperfections in the control lines as well as cross-talk between them will inevitably remain and this problem will be exacerbated by physical crowding as attempts are made to further scale 2D designs.

In this article, a systematic calibration procedure to reduce coherent control error in an all-microwave two-qubit gate is outlined in detail.  This approach allows the reduction of unwanted single-qubit and two-qubit rotations that arise due to large cross-talk and transient behaviour of the control lines, which may for example be caused by line dispersion or reflections at imperfectly matched interfaces in the control wiring. We show that in a readily extensible 3D-integrated circuit architecture \cite{Rahamim2017} with a direct capacitive coupling between coaxial transmons, one can employ the procedure to perform two-qubit quantum logic. This scenario is markedly different from that employing a mediating resonator for the coupling between qubits \cite{Chow2011}, since the control-line cross-talk easily dominates the desired cross-resonance interaction. The systematic approach to calibration presented here is likely to be applicable broadly in two-qubit quantum control, regardless of platform.

The procedure is divided into two major parts. First, a calibration of the applied microwave tones is carried out, using the Hamiltonian tomography technique presented in Ref.~\cite{Sheldon2016}. This is used to estimate cross-talk cancellation parameters without concern for potential transient errors.
Secondly, the minimization of the transient errors is addressed specifically using a tomographic technique based on repetition of the gate that we call ``repeated-gate tomography". We map the dominant two-qubit coherent errors to specific microwave control parameters using spin echoes. The entire procedure is validated on a device consisting of two off-resonant, statically-coupled coaxial transmons, enabling a (CNOT-equivalent) $\hat{ZX}_{-\pi/2}$ gate to be performed with a fidelity of $97.0(7)\%$.

\section{\label{sec:twocoaxmon}Experimental Setup}

\begin{figure*}[!htbp]
\includegraphics[width=1.0\textwidth]{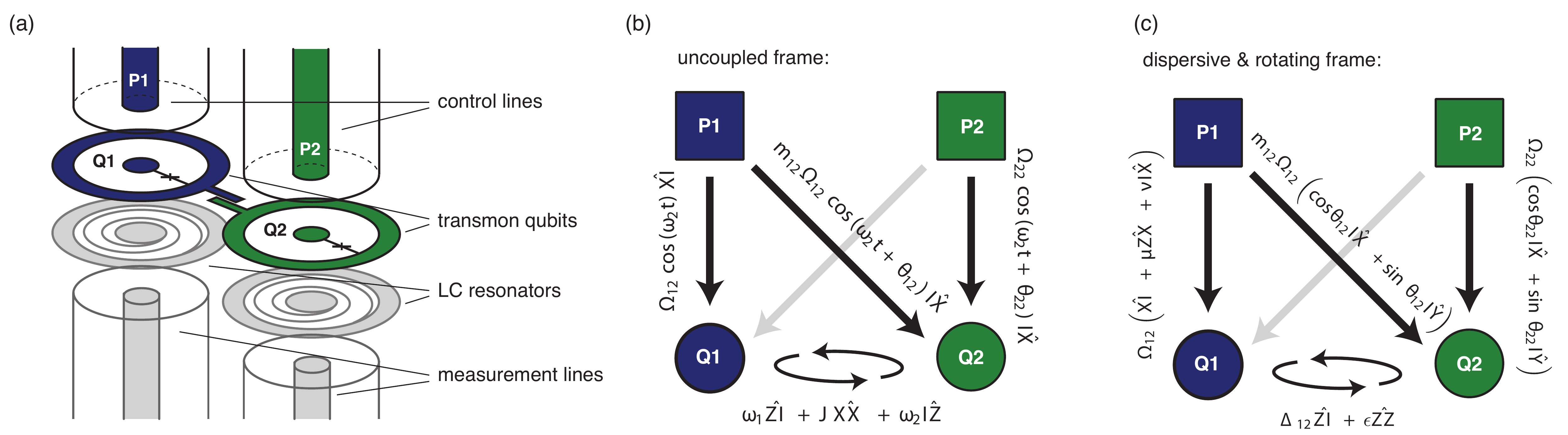}
\caption{\textbf{Device and Hamiltonian Schematic.} (a) Sketch of the coupled two-coaxial-transmon device with associated control lines and readout resonators. (b) Schematic of full Hamiltonian with drive terms due to application of a cross-resonant drive to Q1 at the transition frequency of Q2 (P1$\rightarrow$Q1), a reduced direct application of this drive applied to Q2 due to classical cross-talk with relative amplitude $m_{12}$ and phase $\theta_{12}$ (P1$\rightarrow$Q2), and an on-resonant direct correction drive applied directly to Q2 (P2$\rightarrow$Q2). (c) Transformation of the Hamiltonian terms as in (b) into the frame doubly-rotating with the drives, in the dispersive, low-power limit, $J,\Omega_{12}\ll\Delta_{12}$.
}
\label{fig:2coaxmon}
\end{figure*}

A sketch of the experimental device is shown in \figref{fig:2coaxmon} (a). 
It is a two-qubit version of the coaxial circuit QED architecture presented in \cite{Rahamim2017}. Coaxially-shaped transmon qubits (`coaxmons') lying in one plane on the upper surface of a substrate are capacitively coupled to individual readout resonators on the lower surface of the substrate.  Coaxial drive lines for control of both the transmon and resonator are brought in through the sample holder perpendicular to the substrate surfaces, capacitively coupling to these quantum components. Each qubit, readout resonator and associated control wiring is confined to a cylindrical volume with axis running perpendicular to the substrate surface, producing an architecture which can be extended to larger two-dimensional grids of qubits.

The cross resonance gate is implemented in this architecture through the addition of a small static coupling $J$ between the transmons, achieved here by adding a capacitor between their outer electrodes. We choose this simple direct capacitive coupling instead of employing a mediating resonator in order to minimize the complexity and the number of degrees of freedom in our quantum circuit. Although this may initially appear likely to present a cross-talk challenge due to the proximity of the qubits, we show here that this can be effectively eliminated through good calibration.

Driving one qubit (Q1) at the first transition frequency of the other qubit (Q2) activates an interaction between them that takes place with a rate proportional to the amplitude, $\Omega_{12}$, of this cross-resonant drive \cite{Rigetti2010}. However, other undesired effects are also caused by the presence of this drive. Firstly, the `control' qubit (Q1) under direct drive will be driven off-resonantly. Secondly, due to the presence of higher levels of the transmon, the `target' qubit (Q2) will be directly driven on-resonance, an effect that will be referred to within this article as \emph{quantum cross-talk}.  Thirdly, any stray direct coupling of the control line of the control qubit (P1) to the target qubit (Q2) will cause additional on-resonant driving, an effect referred to here as \emph{classical cross-talk}. Finally, there will be an always-on two-qubit cross-Kerr interaction as the result of the fixed static dispersive coupling, an effect which is ideally minimized by the choice of the circuit parameters (specifically the detuning of the qubit transitions and the control qubits anharmonicity).

The full Hamiltonian of the system, including drive terms, is shown pictorially in \figref{fig:2coaxmon} (b), where $m_{12}$ and $\theta_{12}$ are the relative amplitude and phase of the classical cross-talk term relative to the cross resonance drive, and $\Omega_{22}$ and $\theta_{22}$ are amplitude and phase of a direct correction drive from the port P2 to the qubit Q2 (\emph{cancellation tone}).  

In the frame in which both qubits rotate along with the drive, and in the limit where $J, \Omega_{12} \ll \Delta_{12}$ (where $\Delta_{12} = \omega_1 - \omega_2$ is the detuning between the two transmons), the effective Hamiltonian takes the form,

\begin{align}
2\mathcal{\hat{H}}/\hbar = ~ & \Omega_{ZX} \hat{ZX} + \Omega_{ZY} \hat{ZY} + \Omega_{IX} \hat{IX} + \Omega_{IY} \hat{IY} \nonumber \\& + \Delta_{12} \hat{ZI} + \Omega_{XI}\hat{XI} + \Omega_{YI} \hat{YI} + \epsilon \hat{ZZ} \, ,
\end{align} 

where $\hat{ZX}=\hat{\sigma}_z\otimes\hat{\sigma_x}$ etc. The relevant contributions to each of the rates $\Omega$ are shown in the sketch in \figref{fig:2coaxmon} (c). In the figure, $\mu$ is the cross resonance factor, $\nu$ is the quantum cross-talk factor and $\epsilon$ is the cross-Kerr interaction factor.

The off-resonant control qubit drive terms are defined as  $\Omega_{XI}=\Omega_{12}\cos(\theta_{12})$ and $\Omega_{YI}=\Omega_{12}\sin{\theta_{12}}$.  They can be neglected when the drive amplitude $\Omega_{12} \ll \Delta_{12}$.

The two qubit cross-resonance interaction terms $\Omega_{ZX}=\mu\Omega_{12}\cos{\theta_{12}}$ and $\Omega_{ZY}=\mu\Omega_{12}\sin{\theta_{12}}$ are similarly dependent on the cross-resonance drive amplitude and phase, but also the cross-resonance drive factor $\mu$.  The single qubit rotations on the target $\Omega_{IX} = \Omega_{12} \nu \cos\theta_{12} + \Omega_{12} m_{12} \cos(\theta_{12}+\phi)$ and $\Omega_{IY} = \Omega_{12} \nu \sin\theta_{12} + \Omega_{12} m_{12} \sin(\theta_{12}+\phi)$ are given by the sum of the contribution from the quantum cross-talk characterized by the factor $\nu$ and the classical cross-talk characterized by the factor $m_{12}$ and relative phase $\theta_{12}$.

The terms $\epsilon$, $\mu$ and $\nu$ can be  derived starting with the uncoupled Hamiltonian of the two transmons, finding the transformation that diagonalizes this Hamiltonian and then transforming a drive on the control qubit in the initial non-diagonal Hamiltonian into the new diagonal frame.  A detail of the theory behind a perturbative approach to this is presented in \cite{Richer2013, Julich2013}.  The resulting expressions considering the first three energy levels only of each transmon are

\begin{subequations}
\begin{align}
\mu =& -\frac{J}{\Delta_{12}} \frac{\alpha_1}{\Delta_{12}+\alpha_1}  \, , \label{mu}\\
\nu =& -\frac{J}{\Delta_{12}} \frac{\Delta_{12}}{\Delta_{12}+\alpha_1} \, ,\\
\epsilon =& J^2\frac{\alpha_1+\alpha_2}{(\Delta_{12}+\alpha_1)(\Delta_{12}-\alpha_2)} \, .
\end{align}
\end{subequations}

Table~\ref{Table} shows the relevant parameters of the two-qubit device used in our experiments.

\begin{table}[!htpb]
\begin{tabular}{r|c|l|l}
Qubit & & Q1 & Q2 \\[0.5ex] 
\hline
Transition frequency & $\omega/2\pi~[\mathrm{GHz}]$ & 6.509 & 5.963 \\
Anharmonicity & $\alpha/2\pi~[\mathrm{MHz}]$ & -300 & -314 \\
Relaxation time & $T_1~[\mathrm{\mu s}]$ & 16.2 & 23.9 \\
Coherence time & $T_2~[\mathrm{\mu s}]$ & 25.1 & 35.2 \\[0.5ex] \hline
Cross-Kerr shift & $\Omega_{ZZ}/2\pi~[\mathrm{MHz}]$ & \multicolumn{2}{c}{-0.33}\\
Qubit-qubit coupling & $J/2\pi~[\mathrm{MHz}]$ & \multicolumn{2}{c}{10.7} \\
\end{tabular}
\caption{\textbf{Summary of device parameters.} Relevant parameters of the statically coupled two coaxial-transmon device used to test the calibration procedure.}
\label{Table}
\end{table}

\section{\label{sec:cross-talk} Cross-talk Cancellation}
\begin{figure*}[!htbp]
  \includegraphics[width=1.0\textwidth]{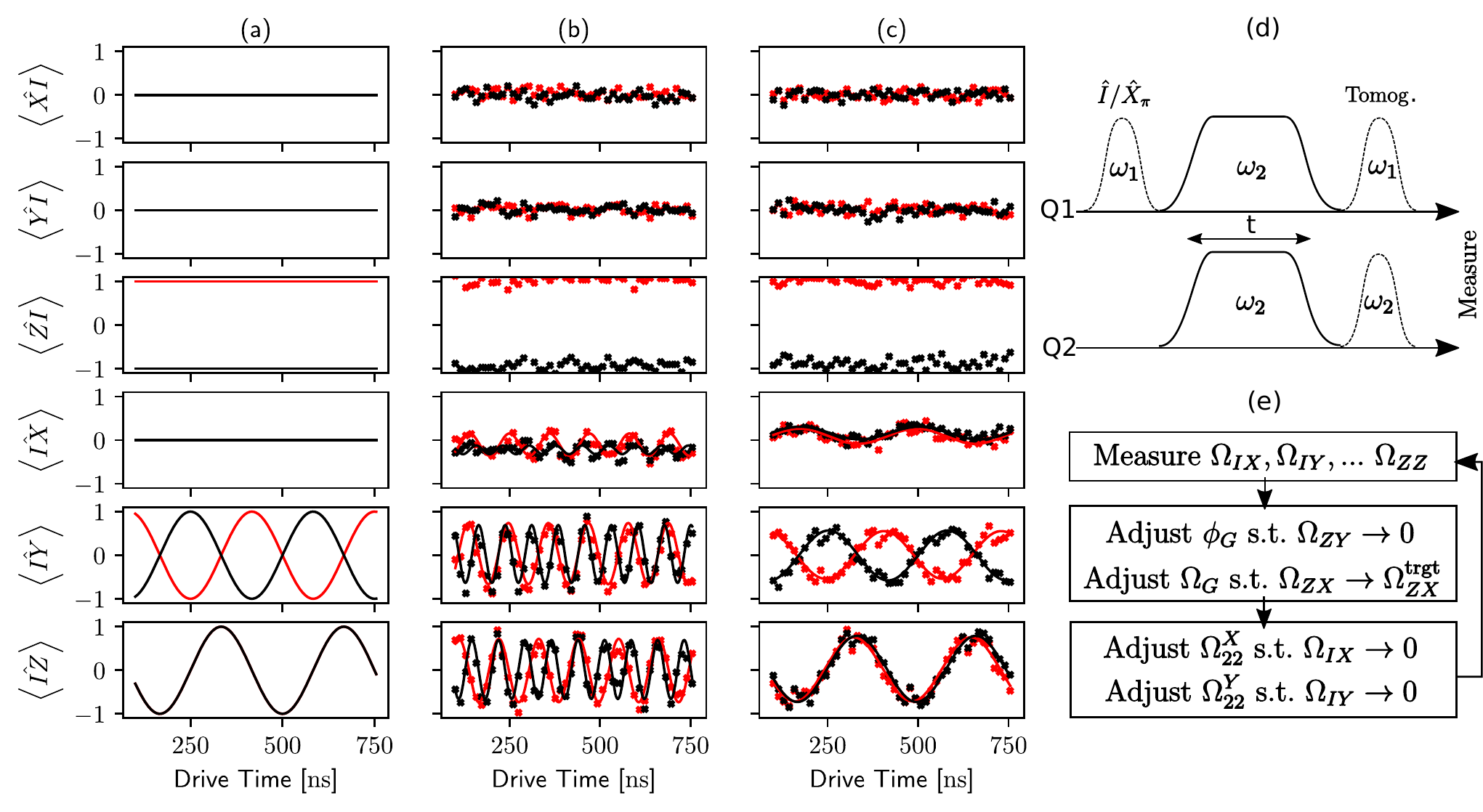}
  \caption{\textbf{Continuous-Wave Cross-Resonance Tuneup with Cross-Talk Cancellation.} Results of simultaneous single qubit tomography on both qubits after a variable length cross-resonant drive pulse has been applied to Q1.  For data shown in red (black), Q1 is initially prepared in state $\ket{0}$ ($\ket{1}$).  (a) Expected results for a Hamiltonian of the form $\mathcal{H}/\hbar=\Omega_{ZX}/2\hat{ZX}$ with $\Omega_{ZX}/2\pi=3.0~\mathrm{~MHz}$. (b) Results using an initial estimate of cross-resonant drive amplitude to achieve this Hamiltonian without application of a cancellation tone.  (c) Results after the amplitude and phase of the cross-resonance tone as well as an introduced cancellation tone have been adjusted to give as close to the desired Hamiltonian as possible.  (d) Sketch of the pulse scheme used in experiments shown in (b) and (c).  (e) Outline of the iterative procedure used to reach the final result in (c).}
  \label{fig:cancellation-calibration}
\end{figure*}
In this section we describe the first part of the calibration process which is used to initially reduce coherent error due to cross-talk.\\
From finite element electromagnetic simulations of the device, we expect to have a classical cross-talk factor $m_{12}$ of order $5\%$. Since this is larger than $\mu = 2.4\%$ (calculated using Equation~\eqref{mu} and Table~\ref{Table}), it is itself enough to cause the single qubit rotation terms $\Omega_{IX}$ and $\Omega_{IY}$ to dominate the dynamics of the system, without considering the level of quantum cross-talk.  Use of echoing schemes alone to reduce this error is not possible, as these terms may be larger than or a similar order to the cross-resonance terms and will not commute with them in general. In order to reduce the $\Omega_{IX}$ and $\Omega_{IY}$ terms, we introduce a cancellation tone directly applied to the target qubit at its own transition frequency with amplitude $\Omega_{22}$ and phase $\theta_{22}$. \\
The optimal values for amplitude and phase of the cancellation tone are obtained using a procedure similar to that presented in Ref.~\cite{Sheldon2016}.  First a cross-resonant microwave pulse of fixed amplitude and varying length $t$ is applied to the control qubit.  Using single-qubit tomography measurements on both qubits, the dynamics of the two individual qubits in time is recorded.  The results for the target qubit are fitted to a fixed-axis, fixed-frequency rotation about the Bloch sphere.
This is possible since the control qubit begins (approximately) in state $\ket{0}$, an eigenstate of $\hat{ZI}$.  This operator commutes with all terms of the Hamiltonian
except $\hat{XI},\hat{YI}$, which we can safely neglect since $\Omega_{XI,YI}\ll \Delta_{12}$. The dynamics are therefore reduced to single qubit dynamics of the target.  Repeating the process with the control qubit initially prepared in state $\ket{1}$ allows $\Omega_{ZX}$, $\Omega_{ZY}$, $\Omega_{IX}$ and $\Omega_{IY}$ to be deduced.  $\Omega_{IZ}$ and $\Omega_{ZZ}$ can also be measured, in theory returning the values $0$ and $\epsilon$ as predicted.

The cancellation tone can then be introduced to the system with a chosen amplitude $\Omega_{22}$ and phase $\theta_{22}$ such that its additive contribution to $\Omega_{IX}$ and $\Omega_{IY}$ are the negative of the values measured previously.  In practice an iterative procedure can be used to find these parameters quickly \cite{Patterson2018}.  Such a process may be useful if non-linearities in the cross-resonance interaction need to be corrected for \cite{Magesan2018arxiv}.

\figref{fig:cancellation-calibration} (b) shows the initial measurement in this process with no cancellation tone applied, and \figref{fig:cancellation-calibration} (c) shows a final measurement with the properly calibrated cancellation tone applied.  The cross-resonant drive phase and amplitude is changed during the procedure to target $\Omega_{ZY}=0$ and $\Omega_{ZX}=\Omega_{CR}^{\text{trgt}}$, where $\Omega_{CR}^{\text{trgt}}$ is a selected target rate at which to drive the cross-resonance interaction. 
A sketch of the pulse scheme used is shown in \figref{fig:cancellation-calibration} (d), while a sketch of the iterative procedure used for the minimization of the cross-talk errors is shown in \figref{fig:cancellation-calibration} (e).

From the cross-resonance interaction rate and the estimated values $\mu = 2.4 \%$ and $\nu = 4.3 \%$, one can calculate the expected contribution of quantum cross-talk to $\Omega_{IX}$ and $\Omega_{IY}$, and subtract it from the total cross-talk.  The remaining classical cross-talk must be the direct result of the drive.  This gives the value of the classical cross-talk parameter $m_{12}=7.1\%$ (corresponding to an isolation of approximately $-23\mathrm{~dB}$).
\subsection{Echo Gate Pulse Scheme}
Using the results shown in \figref{fig:cancellation-calibration} (c), an effective Hamiltonian of the form
\begin{equation}
2\mathcal{\hat{H}}/\hbar \approx \Omega_{CR}^{\text{trgt}} \hat{ZX} + \epsilon \hat{ZZ}
\end{equation}
can be realised.

If $\Omega_{CR}^{\text{trgt}}$ is much larger than $\epsilon$ and any calibration errors, one may expect a pulse of length $t=\Omega_{CR}^{\text{trgt}}/8\pi$ to perform a good approximation of a $\hat{ZX}_{-\pi/2}$ gate.  In practice $\epsilon$ is often non-negligible and small errors in the calibration of the cancellation tone can be present (due to inaccurate calibration or parameter drift).

\begin{figure}[!htbp]
\includegraphics[width=0.45\textwidth]{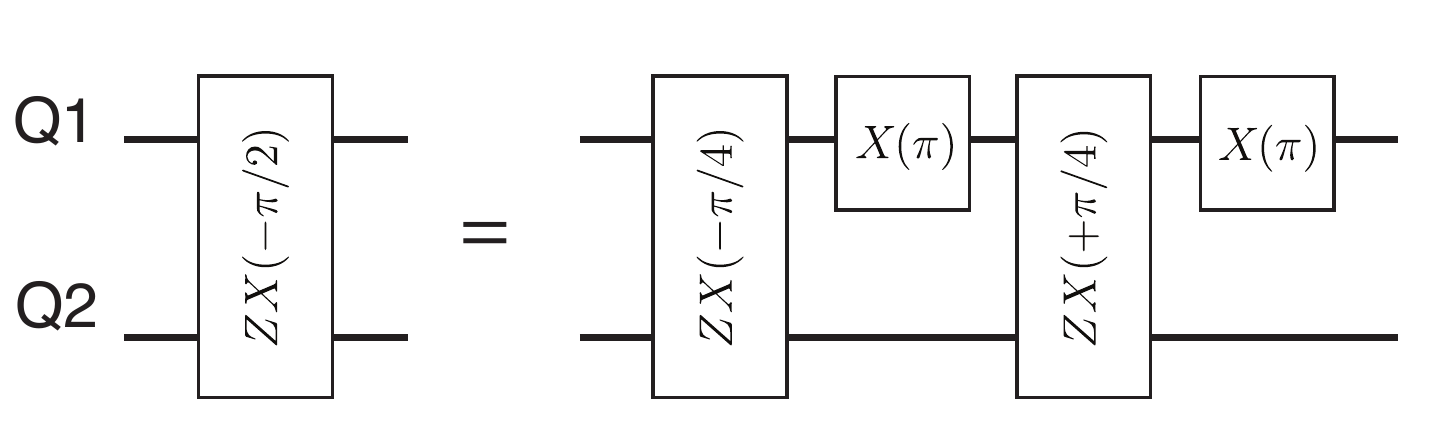}
\caption{\textbf{Echo Gate Scheme.} A $\hat{ZX}_{-\pi/2}$ gate can be formed of two $\hat{X}_\pi$ gates on the control qubit, a $\hat{ZX}_{-\pi/4}$ and a $\hat{ZX}_{+\pi/4}$ gate.  This gate scheme makes the $\hat{ZX}_{-\pi/2}$ gate more resilient to coherent errors common in the $\hat{ZX}_{\pm\pi/4}$ primitives \cite{Sheldon2016}.}
\label{fig:echo-scheme}
\end{figure}

One can reduce these errors in a manner that is robust to slight parameter drift using the echo gate scheme presented in \cite{Sheldon2016}.  This involves performing the cross-resonance gate of length $t$ in two components of length $t/2$, negating the amplitude of the second component and preceding and following it with an $\hat{XI}_{\pi}$ single qubit gate applied to the control qubit.  During the second component, the amplitude of all terms proportional to the drive amplitude ($\Omega_{IX}$, $\Omega_{IY}$, $\Omega_{ZX}$ and $\Omega_{ZY}$) are negated, as are the terms proportional to the control $\hat{Z}$ operator ( $\Omega_{ZX}$, $\Omega_{ZY}$ and $\Omega_{ZZ}$ ).  The only terms not effectively cancelled are $\Omega_{ZX}$ and $\Omega_{ZY}$, the cross-resonance interaction terms, as they are negated twice (see \figref{fig:echo-scheme}).
The effective Hamiltonian across the entire time $t$ of the sequence is then
\begin{equation}
2\hat{H}/\hbar \approx \Omega_{CR}^{\text{trgt}} \hat{ZX} \, .
\end{equation}

This approximation is valid when all terms not commuting with $\hat{ZX}$ in the true Hamiltonian are zero or small in comparison to $\Omega_{ZX}$.  This means that with large values of $m_{12}$ compared to $\mu$, the procedure for cancellation of cross-talk presented previously must be performed and a cancellation tone used before an echo scheme such as the one presented here will begin to help improve the gate quality.  Note that if the largest remaining term in the Hamiltonian after previous calibration is $\Omega_{ZZ}$ as expected, then this echo gate results in an effective Hamiltonian with a smaller error, the leading order term in which will be proportional to $\hat{IY}$ \cite{Patterson2018}.

\section{\label{sec:trans-err}Transient Error Reduction}
In this section we describe the second part of the calibration process which corrects for transient errors using the technique of ``Repeated Gate Tomography''.

\subsection{Repeated Gate Tomography}

\begin{figure*}[!htb]
  \includegraphics{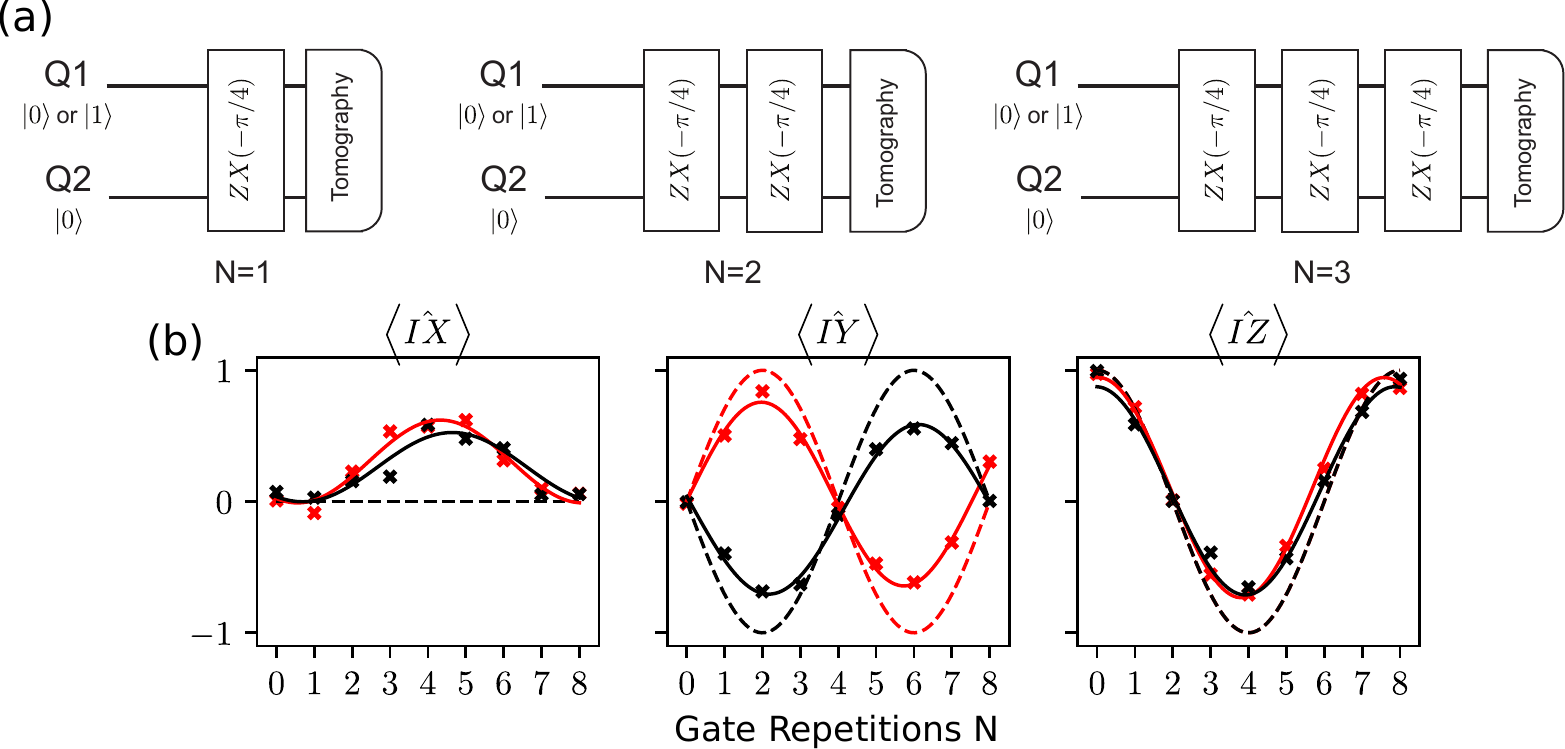}
  \caption{\textbf{Gate characterisation with Repeated-Gate Hamiltonian Tomography.} (a) Pulse scheme used to characterize the $\hat{ZX}_{-\pi/4}$ gate. b) Single qubit tomography results on Q2 after varying repetition numbers of the $\hat{ZX}_{-\pi/4}$ gate are applied to the two qubit system. Red (black) data was collected after Q1 was initially prepared in $\ket{0}$ ($\ket{1})$.  Dashed lines show the ideal case result for a $\hat{ZX}_{-\pi/4}$ gate.}
  \label{fig:gate-tuneup-single}
\end{figure*}

All calibration measurements to this point were performed with pulses of continuously varied lengths, and data has been modelled as a constant rate rotation about a fixed axis.  This means that all Hamiltonian estimates were based upon the extension of a fixed amplitude segment of the pulse for a small time step.  Transient coherent errors due to edge effects of the pulse (such as reflections at the sample holder interface) during the rise and fall of the final gate may be present. If so, their contribution will have an effect on the final gate operation that has not yet been taken into account by the calibration procedure.\\
To alleviate this issue we employ a technique that we call `Repeated Gate Tomography'. Rather than extending the pulse length continuously in time, we instead discretize the pulse length into an integer number of gate repetitions, and repetitions of the gate (including in some cases the entire echo scheme discussed previously) are performed in place of variable length gates. \figref{fig:gate-tuneup-single} (a) shows a sketch of the pulse scheme used.\\

In \figref{fig:gate-tuneup-single} (b), we show the results of single qubit tomography on Q2 after a varying number of repetitions of the $\hat{ZX}_{-\pi/4}$ gate have been applied to the two qubit system, and compare them with the ideal case (dashed lines). To find the effective Hamiltonian we fit the fixed axis, fixed rotation rate model to the data set with a relatively low sampling rate compared to the expected oscillation frequency.  If previous calibration measurements were accurate and the echo scheme is used, the frequency of oscillations in the resulting data will not prevent correct interpretation.  Ascertaining the effective Hamiltonian values (now measured for convenience in units of cycles per gate rather than cycles per second) now allows coherent rotations caused by transient effects during the rise and fall of each pulse to be properly taken into account.\\

\begin{figure*}[!htb]
  \centering
    \includegraphics[width=1.0\textwidth]{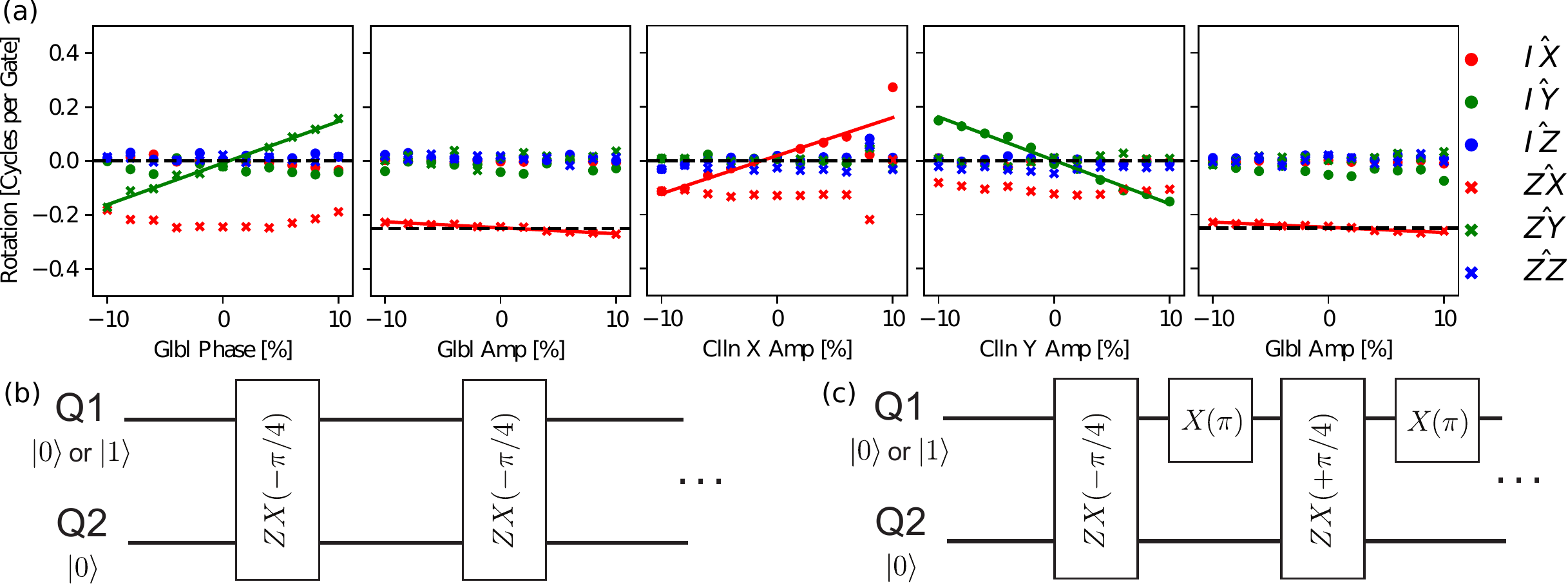}
  \caption{\textbf{Correction of Transient Errors.} (a) Each subplot shows the results of the effective Hamiltonian fit to multiple iterations of an experiment as shown in \figref{fig:gate-tuneup-single} (b), as a physical parameter of the gate implementation is swept in a small range around the previous best estimate.  The global (Glbl) phase (the phase of both pulses with fixed offset) is first swept in a range to correct the $\hat{ZY}$ term.  The global amplitude (the amplitude of both pulses with fixed ratio) is then swept to correct $\hat{ZX}$ to $-1/4\mathrm{~Cycles~per ~Gate}.$ Next the two quadratures of the cancellation tone (Clln X Amp and Clln Y Amp) are swept individually to reduce the remaining $\hat{IX}$ and $\hat{IY}$ terms to zero, before a final re-correction of the $\hat{ZX}$ term. (b) Gate scheme used to deduce the effective Hamiltonian for each value of the Clln X Amp and Clln Y Amp parameters in (a).  The echo scheme is not used here as it reduces the error in $\hat{IX}$ and $\hat{IY}$ to which these parameters are fit. (c) Gate scheme used to deduce the effective Hamiltonian for each value of the Glbl Phase and Glbl Amp parameters.  The echo scheme can be used here as it does not affect the errors in $\hat{ZX}$ or $\hat{ZY}$ to which these parameters are fit.}
  \label{fig:gate-tuneup}
\end{figure*}

\figref{fig:gate-tuneup} shows the extracted effective Hamiltonian from many repetitions of the experiment in \figref{fig:gate-tuneup-single} as various parameters of the gate are varied in a small range around the previous best estimate.

The parameters are varied in the following order. 
First the `Global Phase' $\phi_G$ is adjusted, meaning that the phase of both the cross-resonant $\theta_{12}$ and cancellation $\theta_{22}$ drives are offset from their previous values denoted $\theta_{12}'$ and $\theta_{22}'$ respectively,
\begin{equation}
\theta_{12} = \theta_{12}' + \phi_G,~\theta_{22} = \theta_{22}' + \phi_G \, .
\end{equation}
$\Omega_{ZY}$ has a sinusoidal dependence on $\phi_G$, and so if the previous estimate was good then $\Omega_{ZY}$ will be approximately linearly dependent on $\phi_G$.  The estimate for the global phase parameter is therefore improved by fitting the extracted $\Omega_{ZY}$ values to a linear model and extracting the zero crossing point, as shown in the first inset on \figref{fig:gate-tuneup} (a).

Secondly the `Global Amplitude' $\Omega_G$ is adjusted.  This is the adjustment of the amplitude of the cross-resonant $\Omega_{12}$ and cancellation $\Omega_{22}$ drive amplitudes about their previous values $\Omega_{12}'$ and $\Omega_{22}'$ while keeping a fixed ratio between them,
\begin{equation}
\Omega_{12} = \Omega_G + \Omega_{12}',~\Omega_{22} = \Omega_G \frac{\Omega_{22}'}{\Omega_{12}'} + \Omega_{22}' \, .
\end{equation}
$\Omega_{ZX}$ has a theoretically linear dependence on $\Omega_G$ provided $\Omega_{12}$ remains much less than $\Delta_{12}$.  The estimate for the global amplitude parameter is therefore improved by fitting the $\hat{ZX}$ rotation rate to a linear model and extracting the crossing point with a rotation rate of $-0.25~\mathrm{Cycles~per~Gate}$, as shown in the second inset in \figref{fig:gate-tuneup} (a).

The in-phase and quadrature-phase components of the cancellation tone $\Omega_{22}^X = \Omega_{22}\cos{\theta_{22}}$ and $\Omega_{22}^Y = \Omega_{22}\sin{\theta_{12}}$ are now sequentially adjusted, by individually offsetting them within a small fraction of the total cancellation tone amplitude.

The measured $\Omega_{IX}$ and $\Omega_{IY}$ Hamiltonian terms are then fitted to linear models in the two cases and the relevant parameter updated to the value of the zero-crossing point.  In order to calibrate these parameters, the echo scheme must not be used so that measured single qubit rotations of the target qubit alter with cancellation tone in a simple manner.  During these calibration stages, therefore, a single microwave pulse of length $t/2$ is applied to both control and target (an approximate $\hat{ZX}_{-\pi/4}$ gate) and is repeated a variable number of times.  After calibration of these parameters, $\Omega_G$ is again adjusted to correct the total rotation angle after changes to the other parameters.

The adjustments made in the case shown in \figref{fig:gate-tuneup} (a) appear slight, yet from the simple linear model we predict them to have increased the final gate fidelity by $1.1\%$.

\subsection{Final Results}

Process tomography \cite{nielsen_chuang_2010} and interleaved randomized benchmarking \cite{Corcoles2012} were used to assess the fidelity of gates optimised using the procedure presented above.  The entire procedure including benchmarking can be performed for different gate lengths to find the optimal value for $\Omega_{CR}^{\text{trgt}}$.

\begin{figure*}[!ht]
  \centering
      \includegraphics[width=0.9\textwidth]{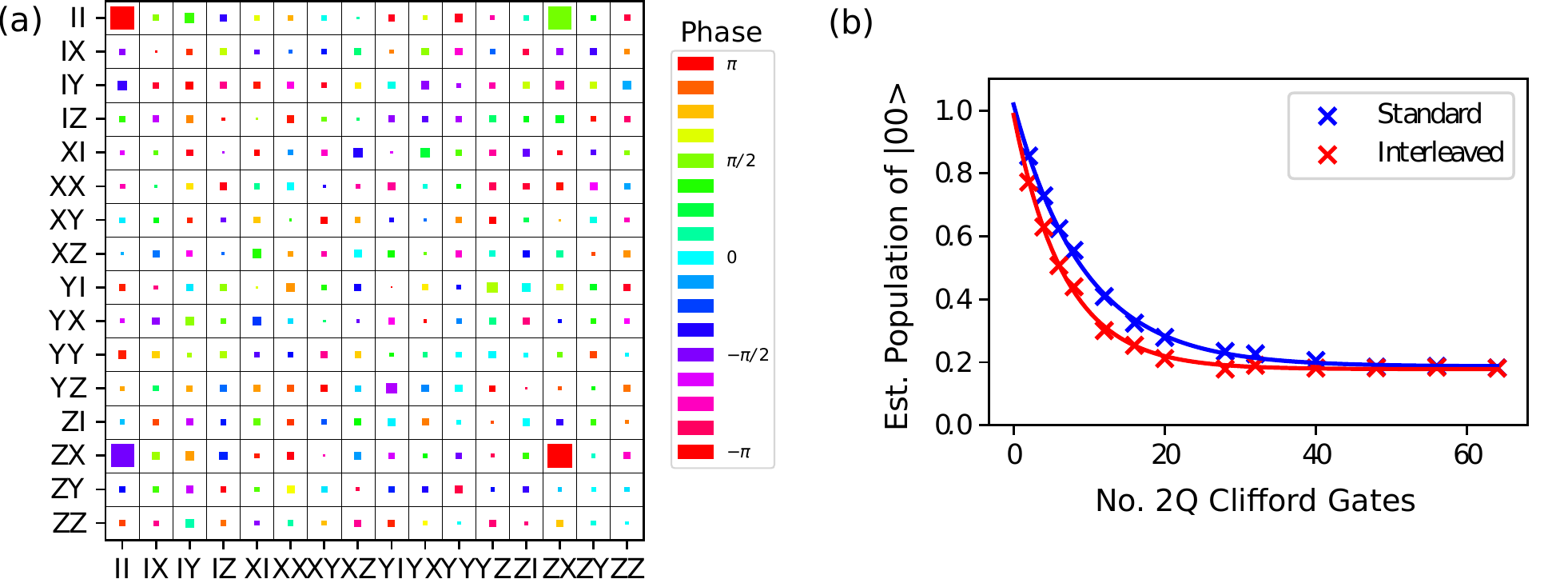}
  \caption{\textbf{Characterisation and Benchmarking of Optimised Gate.} The standard techniques of Quantum Process Tomography and Interleaved Randomized Benchmarking are performed on the optimised gate.  (a) Process tomogram (displayed matrix is $\chi$, the Kraus decomposition of the transfer matrix) provides qualitative confirmation that a $\hat{ZX}_{-\pi/2}$ gate was applied to the system. (b) Results of interleaved randomized benchmarking show that the gate has a fidelity of $97.0\%\pm0.7\%$. The error given is a 90\% confidence interval.}
  \label{fig:final-bm}
\end{figure*}

After this further optimization process, on this particular device a value of $\Omega_{CR}^{\text{trgt}}/2\pi=3.0\mathrm{~MHz}$ allowed a $\hat{ZX}_{-\pi/2}$ gate to be performed with a fidelity of $F=97.0(7)\%$, as measured with interleaved randomized benchmarking as shown in \figref{fig:final-bm} (b).  
The error given is a 90\% confidence interval.  The behaviour of the gate can also be qualitatively evaluated by inspection of the process tomography plot shown in \figref{fig:final-bm} (a).

The predicted fidelity limit for a gate of this length is $98.9\%$ when taking into account both decoherence and single qubit gate error, an error-rate approximately $2.7$ times smaller than that seen here.  There are two likely sources of this discrepancy.  A remaining coherent single qubit error ($\hat{IY}$, resulting from the echoed out $\hat{ZZ}$ interaction) could be corrected to further increase the fidelity. It would also be preferable to reduce $\epsilon$ in future designs. Leakage out of the computational subspace to higher transmon energy levels may also contribute to the gate error.

Simple repetition of this procedure with different target cross-resonance rates in order to search for the optimum gate length can take several hours, largely due to the time required to perform randomized benchmarking. This problem could be mitigated by development of an error model to allow the optimum value of $\Omega_{CR}^{\text{trgt}}$ to be predicted.  Alternatively, full numerical optimisation of the cross-resonance drive pulse could be considered \cite{Allen17, Kirchhoff18}.

\section{\label{sec:conclusion}Conclusion}
In conclusion, we have presented a method for the calibration of the cross-resonance two-qubit gate to reduce error caused by classical cross-talk and imperfectly matched control lines. The method has been validated using a two-qubit implementation of coaxial circuit QED \cite{Rahamim2017} with directly coupled transmons. The procedure outlined here is applicable to other physical systems that use a driven cross resonance two-qubit interaction, and can serve as a general guide for mitigating unwanted coherent errors in quantum processors.


\section*{Acknowledgments}
This work has received funding from the UK Engineering and Physical Sciences Research Council under grant nos.~EP/J013501/1, EP/M013243/1 and EP/N015118/1, and from Oxford Quantum Circuits Limited. A.P. acknowledges Oxford Instruments Nanoscience, M. M. acknowledges the Stiftung der Deutschen Wirtschaft (sdw) and T.T. acknowledges the Masason Foundation and the Nakajima Foundation for financial support.

\bibliography{library}

\end{document}